\documentclass[twocolumn]{aastex631_cranmer2023}
\usepackage{hyperref}
\usepackage{times}

\shorttitle{MHD Mode Conversion in the Solar Corona}
\shortauthors{S.\  R.\  Cranmer \& M.\  E.\  Molnar}

\begin{document}

\title{Magnetohydrodynamic Mode Conversion in the Solar Corona:
Insights from Fresnel-like Models of Waves at Sharp Interfaces}

\correspondingauthor{Steven R. Cranmer}
\author[0000-0002-3699-3134]{Steven R. Cranmer}
\affiliation{Department of Astrophysical and Planetary Sciences,
Laboratory for Atmospheric and Space Physics,
University of Colorado, Boulder, CO, USA}

\author[0000-0003-0583-0516]{Momchil E. Molnar}
\affiliation{Department of Astrophysical and Planetary Sciences,
Laboratory for Atmospheric and Space Physics,
University of Colorado, Boulder, CO, USA}
\affiliation{National Solar Observatory, Boulder, CO, USA}
\affiliation{High Altitude Observatory, National Center for Atmospheric
Research, Boulder, CO, USA}

\begin{abstract}
The solar atmosphere is known to contain many different types of
wavelike oscillation.
Waves and other fluctuations (e.g., turbulent eddies) are
believed to be responsible for at least some of the energy
transport and dissipation that heats the corona and accelerates
the solar wind.
Thus, it is important to understand the behavior of magnetohydrodynamic
(MHD) waves as they propagate and evolve in different regions of the
Sun's atmosphere.
In this paper, we investigate how MHD waves can affect the
overall plasma state when they reflect and refract at sharp, planar
interfaces in density.
First, we correct an error in a foundational paper (Stein 1971)
that affects the calculation of wave energy-flux conservation.
Second, we apply this model to reflection-driven MHD turbulence in
the solar wind, where the presence of density fluctuations can enhance
the generation of inward-propagating Alfv\'{e}n waves.
This model reproduces the time-averaged Elsasser imbalance fraction
(i.e., ratio of inward to outward Alfv\'{e}nic power) from several
published numerical simulations.
Lastly, we model how the complex magnetic field threading the
transition region between the chromosphere and corona helps convert
a fraction of upward-propagating Alfv\'{e}n waves into
fast-mode and slow-mode MHD waves.
These magnetosonic waves dissipate in a narrow region around the
transition region and produce a sharp peak in the heating rate.
This newly found source of heating sometimes exceeds the expected
heating rate from Alfv\'{e}nic turbulence by an order of magnitude.
It may explain why some earlier models seemed to require an additional
ad-hoc heat source at this location.
\end{abstract}


\section{Introduction}
\label{sec:intro}

The Sun's photosphere, chromosphere, and corona are highly dynamic,
exhibiting stochastic and intermittent variability across many
orders of magnitude in space and time.
These fluctuations have long been suspected to be important
contributors to how the plasma is heated
\cite[see, e.g.,][]{Asch06,PD12,Fl15,VD20}.
Many of the proposed conceptual mechanisms involve oscillating
magnetohydrodynamic (MHD) waves.
Some propose that convective motions drive oscillations at the solar
surface that propagate up to larger heights and then dissipate
to provide coronal heating \citep{A47,Sp81,MU02,CvB05,Js23}.
Others suggest that the evolving magnetic field triggers isolated
bursts of magnetic reconnection that generate wave activity throughout
the solar atmosphere \citep{Ho90,AM92,Ki10,Ly14}.

In addition to the relatively straightforward mechanisms described above,
there are likely to be many other indirect ways for MHD waves to be
generated and to evolve (i.e., to transform their properties as they
propagate from one place to another in the solar atmosphere).
The literature on both linear and nonlinear {\em mode conversion}
is vast, and we can only provide pointers to a small fraction of it
for further study.
When the magnetic field strength and other background plasma properties
(e.g., density and temperature) vary with position, it allows waves of
one type to transform into another
\citep[see, e.g.,][]{LR86,Bg02,MH07,CG08,Tr17,Cy22,Hu22}.
In addition, when waves encounter inhomogeneities such as velocity
shears, there can arise both linear \citep{Bv78,Kg99,Kg07,Kg13,HK12}
and nonlinear \citep{Nk98,Bc22} channels to convert some waves into
different modes.
In many realistic inhomogeneous environments, the modes
themselves become ``mixed'' and the usual separation into ideal fast,
slow, and Alfv\'{e}n modes is not possible \citep[e.g.,][]{Goo19}.
When MHD waves propagate through random or stochastic background
conditions, it can trigger phenomena such as resonances,
instabilities, and turbulent cascade
\citep{Va71,Va74,Lou86,LZ87,Mu01,Yu15,Sr21,Mt23}.
Lastly, when wave amplitudes become large, there can arise various
varieties of shock, soliton, jet, and vortex phenomena that render
meaningless the dividing lines between textbook ``normal modes'' of
wavelike oscillation \citep[e.g.,][]{VH96,Hs03,RH07,CW15,Sn18}.

Despite the dizzying array of proposed mechanisms of MHD wave mode
conversion, there are a few that we believe have not received sufficient
attention.
Specifically, in this paper we examine some of the consequences of a
model first discussed by \citet{St71}.
That paper studied the consequences for a wave obliquely incident on a
sharp planar boundary at which the plasma density varies abruptly.
In ideal MHD, a wave of one type (i.e., slow-mode, fast-mode, or
Alfv\'{e}nic) produces reflected and transmitted power in waves of
all three types.
In Section~\ref{sec:stein}, we describe this model in detail and
correct an error in one of the equations given by \citet{St71}.
In Section~\ref{sec:alf}, we apply this model to the solar wind in order
to better understand how density fluctuations can enhance the large-scale
reflection of imbalanced Alfv\'{e}nic turbulence.
In Section~\ref{sec:TRopen}, we simulate the Sun's complex
``magnetic carpet'' in order to see how Alfv\'{e}n waves incident on
the sharp transition region (TR) will naturally transfer some of their
energy to both upward and downward propagating fast/slow-mode
magnetosonic waves.
These compressive waves dissipate very efficiently at the TR, so they
appear to provide an added ``pulse'' of coronal heating that
has not been considered before.
Lastly, in Section~\ref{sec:conc}, we conclude by summarizing these
results, discussing possible future improvements, and suggesting
some additional applications of the \citet{St71} theory.

\section{Reflection and Transmission of MHD Waves}
\label{sec:stein}

Here we describe the setup of an idealized MHD system containing a
planar discontinuity in the background plasma parameters.
We often follow the notation of \citet{St71}, but we also correct an
error in one of Stein's equations that led to the normal component
of the total wave energy flux to not be conserved.
We also take inspiration from additional studies of waves at sharp
interfaces such as \citet{Va90}, \citet{Te11}, and \citet{Vi18}, and
we anticipate that many of the results discussed below
will also be valid for environments where the interface has a
{\em finite} thickness smaller than the other relevant length-scales
(e.g., scale heights and wavelengths) of the medium.

\begin{figure}[!t]
\epsscale{1.00}
\plotone{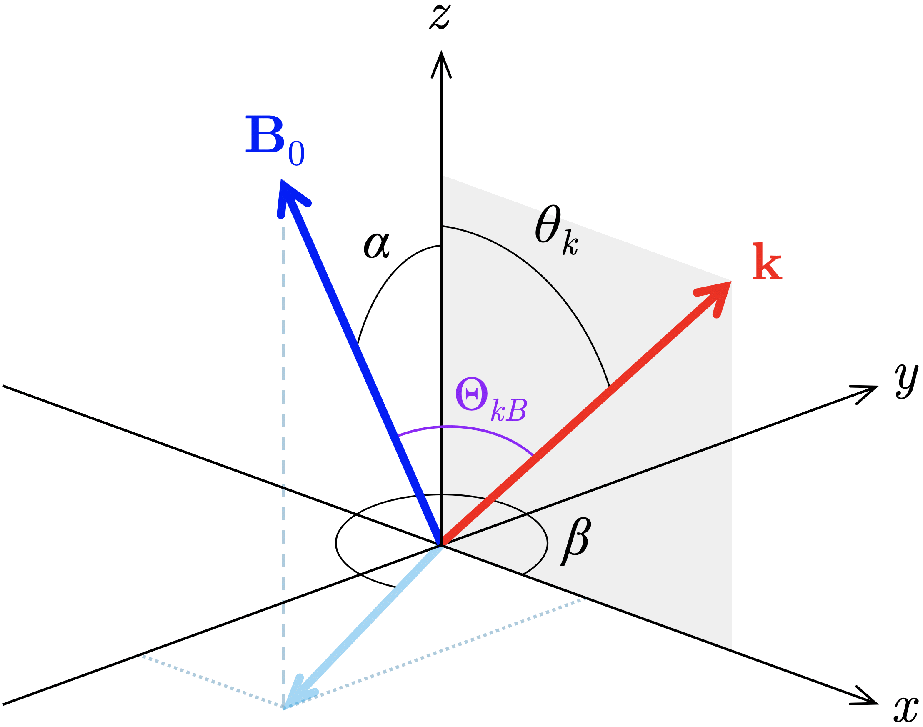}
\vspace*{0.05in}
\caption{Coordinate system and vectors for the idealized MHD system of
wave reflection and transmission.
Wavenumber $\bf k$ (red) always remains in the $x$-$z$ plane, and
it is offset from the $z$-axis by a polar angle $\theta_k$.
Spherical coordinate angles $\alpha$ and $\beta$ describe the
magnetic field vector (blue).
\label{fig01}}
\end{figure}

We define coordinates such that the discontinuity occurs in the plane
described by $z=0$, and in general the mass density $\rho$,
temperature $T$, and magnetic field strength $B_0$ can be different
in the two homogenous half-spaces defined by $z<0$ and $z>0$.
In this work, we choose to keep the both the magnetic field and
the background gas pressure ($P_0 \propto \rho T$) constant across
the discontinuity.
Thus, we specify $\rho_1$ and $T_1$ for $z<0$,
and $\rho_2$ and $T_2$ for $z>0$, with
\begin{equation}
  \rho_1 / \rho_2 \, = \, T_2 / T_1 \,\,\, .
\end{equation}
Figure \ref{fig01} illustrates various quantities in the Cartesian
coordinate system defined by the discontinuity in the $x$-$y$ plane.
An incident MHD plane-wave travels up from the $z<0$ region,
encounters the discontinuity, and creates a set of transmitted
($z>0$) and reflected ($z<0$) waves as a response.
Each of those linear waves has its own wavenumber vector $\bf k$,
but they all share the same angular frequency $\omega$.

The angle between $\bf k$ and the magnetic field vector ${\bf B}_0$
is defined as $\Theta_{kB}$, with
\begin{equation}
  \cos\Theta_{kB} \, = \, \sin \theta_k \sin\alpha \cos\beta
  + \cos\theta_k \cos\alpha  \,\, .
\end{equation}
The characteristic speeds of the plasma are defined as
\begin{equation}
  V_{\rm A} \, = \, \frac{B_0}{\sqrt{4\pi\rho_i}}
  \,\,\,\,\,\, \mbox{and} \,\,\,\,\,\,
  c_{s} \, = \, \sqrt{\frac{\gamma P_0}{\rho_i}}
\end{equation}
for the Alfv\'{e}n speed and adiabatic sound speed, respectively,
with $i=1,2$ depending on the region being described.
We always used $\gamma = 5/3$, and we also used the ideal-MHD
versions of the fast, slow, and Alfv\'{e}nic dispersion relations
that were given for this coordinate system in Equations (3)--(7)
of \citet{St71}.

In general, all three types of MHD wave modes may be generated above
and below the discontinuity.
A range of properties must be continuous at the interface, including
the frequency $\omega$ and the in-plane component of the wavevector $k_x$.
The dispersion relations are then solved for $k_z$ on both sides of
the interface.
However, \citet{St71} discussed how some combinations of parameters
lead to one or both of the resulting fast-mode waves having a purely
imaginary normal wavenumber $k_z$.
In these cases, the fast-mode wave is evanescent and has zero energy
flux in the direction normal to the discontinuity.
However, this mode still needs to be taken into account when solving
for the amplitudes at the interface.
We used a complex-variable form of $LU$ decomposition \citep{Tu48,An99}
to solve the associated 6~$\times$~6 linear system for the amplitudes.

Once the dimensionless amplitudes $A$ are known for the reflected
and transmitted wave modes, one can compute $F_z$, the component of
energy flux normal to the discontinuity, for each mode.
Strictly speaking, all reflected modes must have $F_z < 0$ and
both the incident and transmitted modes must have $F_z > 0$.
Note that Equation (B4) of \citet{St71} defined the transverse
velocity amplitude of Alfv\'{e}n waves as
\begin{equation}
  u_{\perp} \, = \,
  A V_{\rm A} \left| \hat{\bf e}_k \times \hat{\bf e}_B \right|
  \, = \,
  A V_{\rm A} \sin\Theta_{kB} \,\, ,
\end{equation}
where $\hat{\bf e}_k$ and $\hat{\bf e}_B$ are unit vectors parallel
to ${\bf k}$ and ${\bf B}_0$.
However, Equation (B16) of \citet{St71} defined the $z$-component
of the wave energy flux using the definition $A = u_{\perp}/V_{\rm A}$.
Using the proper definition above, the corrected expression for
the Alfv\'{e}n-wave flux becomes
\begin{equation}
  F_z \, = \, A^2 \rho_0 V_{\rm A}^3 \,
  \cos\alpha \, \sin^2 \Theta_{kB} \,
  \frac{\cos\Theta_{kB}}{| \cos\Theta_{kB} |} \,\, .
\end{equation}
The corresponding expressions for fast and slow magnetosonic waves
given by \citet{St71} are correct.

Lastly, we define the reflection and transmission coefficients 
\begin{equation}
  {\cal R}_{ij} \, = \, \left| \frac{F_{z,i} \mbox{(reflected)}}
  {F_{z,j} \mbox{(incident)}} \right|
  \,\,\,\,\,\, \mbox{and} \,\,\,\,\,\,
  {\cal T}_{ij} \, = \, \left| \frac{F_{z,i} \mbox{(transmitted)}}
  {F_{z,j} \mbox{(incident)}} \right|
\end{equation}
where, in general, the subscript $i$ can be A, F, or S (for any of the
three modes generated at the interface) and the subscript $j$ (for the
incident wave) is chosen as only one of either A, F, or S.
We always verified that the numerical results satisfy
\begin{equation}
  \sum_i \left( {\cal R}_{ij} + {\cal T}_{ij} \right) \, = \, 1 \,\, ,
\end{equation}
and that the results are independent of the numerical value chosen for
the wave frequency $\omega$.
Throughout the remainder of this paper, we consider only the case of
the incident waves being Alfv\'{e}nic (i.e., $j = {\rm A}$),
and we discuss other scenarios in Section~\ref{sec:conc}.

\section{Enhanced Reflection of Alfv\'{e}nic Turbulence}
\label{sec:alf}

As a first application of the idealized system described above, we
would like to better understand the results of numerical simulations
that contain both counter-propagating Alfv\'{e}n waves and
stochastic density fluctuations.
Investigations of MHD turbulence in the solar corona and solar wind
have found that there is more Alfv\'{e}n-wave reflection when the
density fluctuations are stronger \citep[see, e.g.,][]{vB16,vB17}.
Thus, we explore whether it is possible to model this behavior as a
series of reflections and transmissions through multiple discontinuities.

\subsection{Alfv\'{e}n Waves at a Single Interface}
\label{sec:alf:one}

When the angle $\beta = 0$, the background magnetic field vector
${\bf B}_0$ lies in the $x$-$z$ plane defined by the wavevector
${\bf k}$ and the normal to the discontinuity.
In that case, the 6~$\times$~6 matrix decouples into a
2~$\times$~2 matrix for Alfv\'{e}n waves and a 
4~$\times$~4 matrix for the fast and slow modes.
\citet{Va90} explored this limiting case and discussed the
straightforward analytic solution for Alfv\'{e}n waves,
\begin{equation}
  {\cal R}_{\rm AA} \, = \, \left( \frac{1 - \Delta}{1 + \Delta}
  \right)^2 \,\, ,
  \label{eq:RAA1}
\end{equation}
where $\Delta = \sqrt{\rho_2 / \rho_1}$.
Interestingly, this result is independent of the values of
$\alpha$ and $\theta_k$, as well as independent of the background
magnetic field strength.
Also, it is the same even if the incident wave goes in the other
direction through the interface (i.e., if the
density jump $\Delta$ is replaced by $1/\Delta$).

\subsection{Alfv\'{e}n Waves at a Multiple Interfaces}
\label{sec:alf:multi}

\begin{figure}[!t]
\vspace*{0.05in}
\epsscale{1.00}
\plotone{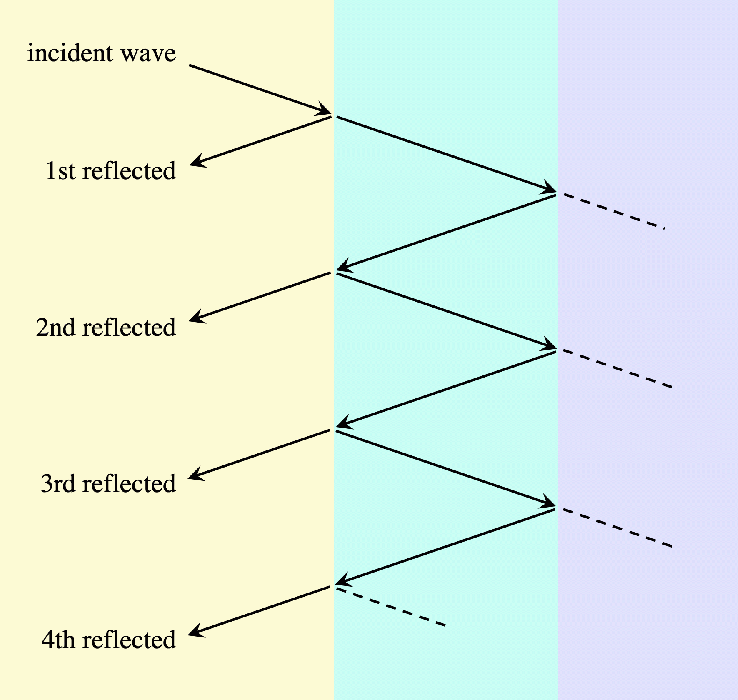}
\caption{Illustration of how reflection from two parallel interfaces
results in an infinite series of reflected components.
\label{fig02}}
\end{figure}

For a system containing multiple interfaces (all assumed to be
mutually parallel planes with identical magnitudes $\Delta$),
it is possible to determine the net 
effect of Alfv\'{e}n-wave transmission and reflection through them.
Even for just two interfaces, the total reflection coefficient is the
result of an infinite series of reflected waves.
Figure~\ref{fig02} illustrates this infinite series.
Specifically, the first reflection is immediate.
The second is the result of one transmission, then one reflection (at
the second interface), then another transmission back toward the source.
The third is the result of one transmission, then three ``internal bounce''
reflections, then another transmission back to the source.
The fourth is the result of one transmission, then five reflections, then
another transmission, and so on.
Using the shorthand notation ${\cal R}_1$ for the reflection
coefficient given in Equation (\ref{eq:RAA1}), the total reflection
coefficient for two interfaces is given by
\begin{equation}
  {\cal R}_2 \, = \, {\cal R}_1 + {\cal T}_1^2 {\cal R}_1 \left(
  \, 1 + {\cal R}_1^2 + {\cal R}_1^4 + {\cal R}_1^6 +
  \cdots \, \right) \,\, .
\end{equation}
Note that the following infinite series converges,
\begin{equation}
  \sum_{n=0}^{\infty} {\cal R}_1^{2n} \, = \, \frac{1}{1 - {\cal R}_1^2}
\end{equation}
as long as $|{\cal R}_1| < 1$.
Thus, given that ${\cal T}_1 = 1 - {\cal R}_1$ for the interaction
at a single discontinuity, the entire expression simplifies to
\begin{equation}
  {\cal R}_2 \, = \, \frac{2{\cal R}_1}{1 + {\cal R}_1} \,\, .
\end{equation}
This kind of calculation has been generalized to $N$ interfaces, with
a general solution given by
\begin{equation}
  {\cal R}_N \, = \, \frac{N{\cal R}_1}{1 + (N-1){\cal R}_1} \,\, .
  \label{eq:RN}
\end{equation}
Note that, in the limit of $N \rightarrow \infty$, the total
reflection coefficient ${\cal R}_N \rightarrow 1$.

These results depend on the multiple interfaces being
effectively ``sharp'' in comparison with the length scales of the
waves (i.e., wavelengths corresponding to components of ${\bf k}$
normal to the interface).
It is also likely that the results depend on the interfaces being
positioned at least one or more wavelengths apart from one another,
rather than being stacked close together.
For additional examples of Alfv\'{e}n-wave reflection and transmission
at multiple interfaces, see \citet{Ho84} and \citet{DP01}.

\subsection{Application to Compressible Fluctuations}
\label{sec:alf:app}

To apply these concepts to simulations of heliospheric turbulence,
we need to model a given spatial distribution of density fluctuations
as a series of sharp jumps.
Thus, given the statistical properties of the fluctuations, the goal
is to compute the magnitude $\Delta$ for each (presumed identical) jump.
If the fluctuations can be approximated as a periodic waveform, with
a mean density $\rho_0$ and variations that extend to $\pm \Delta \rho$
above and below that mean, then
\begin{equation}
  \Delta \, = \, \sqrt{\frac{\rho_0 + \Delta\rho}{\rho_0 - \Delta\rho}}
\end{equation}
as a representative interface ratio.
However, for specific waveform shapes, the amplitude $\Delta\rho$
relates to the root-mean squared (rms) density---i.e., the square root
of the variance---in different ways.
Using $\delta \rho / \rho_0$ for the ratio of the rms to the mean, we
can write
\begin{equation}
  \Delta \, = \, \sqrt{\frac{1 + (\delta\rho / \rho_0)\sqrt{s}}
  {1 - (\delta\rho / \rho_0)\sqrt{s}}} \,\, ,
\end{equation}
where $s$ is a dimensionless shape factor; see, e.g., Equation~(38)
of \citet{CvB07}.
For a square-wave train, $s=1$.
For a sinusoidal wave, $s=2$.
For a sawtooth or triangular waveform, $s=3$.

\begin{figure}[!t]
\vspace*{0.07in}
\epsscale{1.18}
\plotone{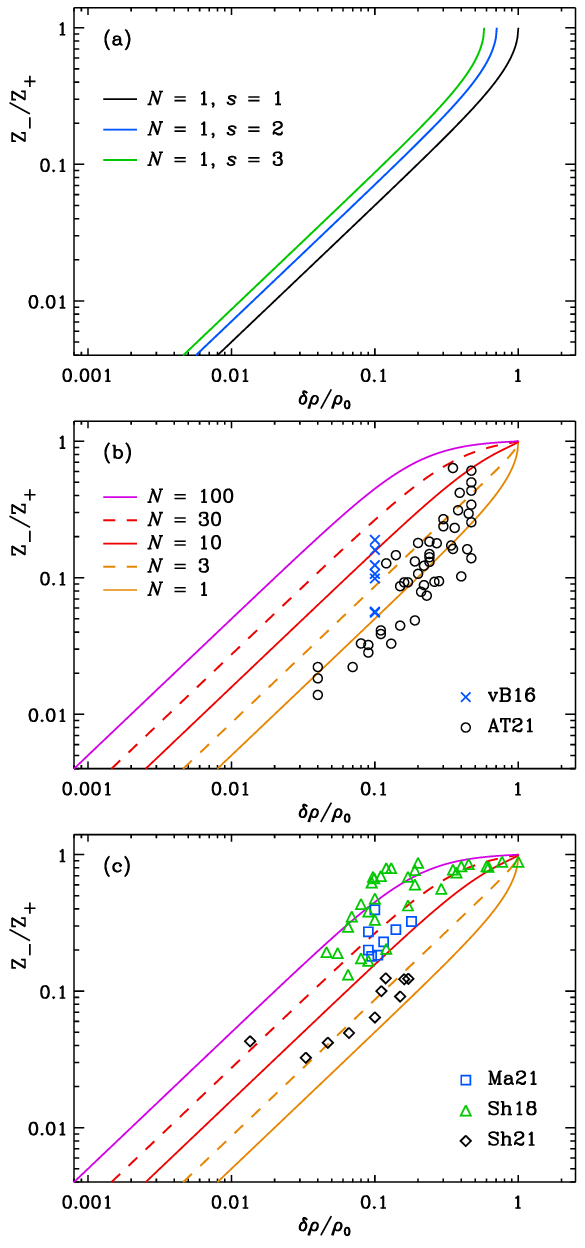}
\caption{Degree of Alfv\'{e}n-wave reflection, plotted as an
amplitude ratio, as a function of a
normalized rms density fluctuation.
(a) For $N=1$, we vary the shape factor $s=1$ (black curve),
$s=2$ (blue curve), $s=3$ (green curve).
(b) For various values of $N$, we compare model results
(see legend for curve colors and types) with data points extracted
from numerical simulations.
(c) Same as (b), but for different simulations.
Reference abbreviations are vB16 \citep{vB16}, AT21 \citep{As21},
Ma21 \citep{Ms21}, Sh18 \citep{Sh18}, and Sh21 \citep{Sh21}.
\label{fig03}}
\end{figure}

Figure \ref{fig03}(a) shows the results of using all three values
of $s$ to predict the reflection from a single interface.
In this case, the plotted quantity is the ratio of inward to
outward \citet{E50} amplitudes,
which is often specified for simulations of imbalanced
MHD turbulence.
Here, we associate this ratio with the reflection coefficient as
\begin{equation}
  \frac{Z_{-}}{Z_{+}} \, = \, \sqrt{{\cal R}_1} \,\, .
\end{equation}
We also note that the Alfv\'{e}n speed appears in the definitions
of $Z_{\pm}$, and we presume that the unperturbed density $\rho_0$
is used in these definitions.
In the limit of weak density fluctuations, this gives
\begin{equation}
  \frac{Z_{-}}{Z_{+}} \, \approx \, \frac{\sqrt{s}}{2}
  \left( \frac{\delta \rho}{\rho_0} \right)  \,\, .
\end{equation}
For simplicity, we use $s=1$ for the models of multiple reflections
discussed below.
In Figures \ref{fig03}(b) and \ref{fig03}(c) we show how the total
amount of Alfv\'{e}n-wave reflection varies for different numbers of
interfaces; i.e., in this case the Elsasser amplitude ratios are
assumed to be equal to ${\cal R}_N^{1/2}$, for various values of $N$.

We compare the multiple-reflection model curves with the outputs of
several different MHD turbulence simulations.
Table~\ref{table01} provides the
base-10 logarithms of both $Z_{-}/Z_{+}$ and $\delta \rho / \rho_0$,
extracted at various heliocentric distances in these simulations.
Specifically,
data points from \citet{vB16} were extracted from their
Figures~2(a) and 4(c),
data points from \citet{As21} were extracted from their
Figures~4, 5, and 6,
data points from \citet{Ms21} were extracted from their
Figures~2(c) and 2(d),
data points from \citet{Sh18} were extracted from their
Figures~2(a) and 2(b),
and data points from \citet{Sh21} were extracted from their
Figures~3 and 4 and Table~1.

For some of the simulations, Table~\ref{table01} gives both the full
Elsasser imbalance ratio (for the model containing the density
fluctuations) as well as a minimum ratio corresponding to only wave
reflection due to large-scale gradients in the smooth background atmosphere;
the latter is denoted $(Z_{-}/Z_{+})_0$ when it is given.
To attempt to isolate the effect of the density fluctuations, the
quantity plotted in Figure \ref{fig03} is
\begin{equation}
  \left( \frac{Z_{-}}{Z_{+}} \right)_{\rm eff} \, = \,
  \sqrt{ \left( \frac{Z_{-}}{Z_{+}} \right)^2 - 
         \left( \frac{Z_{-}}{Z_{+}} \right)_0^2 }  \,\, ,
\end{equation}
but we also note that this is nearly always a very small correction.
Plotting the full ratio would have produced noticeable changes in the
data points shown in Figure \ref{fig03}(b) only for a few of the
points with the lowest values of $Z_{-}/Z_{+}$.

Figure~\ref{fig03} shows some degree of agreement between the modeled
trends and the simulated data.
We note that no other significant correlations were found between the
Elsasser amplitudes and other properties of the turbulence
simulations (say, radial distance from the Sun or background magnetic
field strength).
The agreement seen in Figure~\ref{fig03} allows us to speculate about
best-fitting values for $N$ for each group of data points.
To do this, we minimized a $\chi^2$ difference between the base-10
logarithms of the simulated values of $Z_{-}/Z_{+}$ and a fine grid of
model curves for different values of $N$.
Because we are attempting to model long-time or ensemble averages of an
instrinsically stochastic system, we allow for non-integer values of $N$.

For the combined vB16 and AT21 data in Figure \ref{fig03}(b),
the best-fitting value of $N$ was 1.32.
Note that these simulations contain only incompressible Alfv\'{e}nic
fluctuations that propagate on a background with imposed density
fluctuations.
In general, an order-unity value of $N$ may make sense because
the eddies in strong MHD turbulence do not persist forever like
sinusoidal wavetrains.
They decay, often substantially, over just one to a few
wavelengths or parallel correlation lengths
\citep[see, e.g.,][]{ZM90,GS95,Sch22}.
Thus, any given packet of Alfv\'{e}nic turbulent energy may only live
long enough to interact with one or two of our model density interfaces.

For the data in Figure \ref{fig03}(c), each group was treated separately.
For Sh18, the best-fitting $N \approx 76.7$,
for Ma21, the best-fitting $N \approx 21.5$, and
for Sh21, the best-fitting $N \approx 3.38$.
In contrast to the models shown in panel (b), these models contain
additional nonlinear sources of wavelike density fluctuations---such as
those induced by the parametric decay instability---that propagate
up and down relative to the Alfv\'{e}n waves.
Thus, one may expect to see a larger value of $N$ than in the case
of pure Alfv\'{e}nic turbulence.
Note that the largest effective values of $N$ correspond to the
one-dimensional models of \citet{Sh18}, in which it has been suggested
that wavetrains have longer lifetimes when not allowed to evolve
in the directions transverse to the background magnetic field.

These results can be applied to future models of the solar wind
that contain reflection-driven Alfv\'{e}nic turbulence.
Specifically, by defining a ``floor'' value of $\delta \rho / \rho_0$,
below which we do not expect the solar wind to ever dip below, we could
thus impose a corresponding floor on the Elsasser imbalance ratio.
However, some caution is needed.
For example, depending on the actual radial length-scales of the
density fluctuations, there may only exist a finite number of
effective interfaces between the Sun and some arbitrary radial
distance $r$.
Thus, it may not be appropriate to use the largest values of
$N$ found above (i.e., $N \approx 75$) because there may not be room
for that many interfaces.

\section{The Complex Transition Region}
\label{sec:TRopen}

This paper's second application of the \citet{St71}
reflection/transmission theory is to the sharp transition region (TR)
between the Sun's chromosphere and corona.
There have been multiple studies of how Alfv\'{e}n waves are affected
when they encounter this thin interface
\citep[e.g.,][]{Ho78,Ho81,Wz78,CM95,CvB05,TK21,ST21}.
Many of these studies assumed a relatively simple
geometry for the magnetic field that threads the TR.
However, the properties of wave-mode conversion at a discontinuity
depend sensitively on the magnetic-field angles $\alpha$ and $\beta$
shown in Figure~\ref{fig01}.
Thus, our goal is to quantify the degree of wave reflection,
transmission, and conversion for realistic ensembles of
three-dimensional field-line geometries.

Before describing our model, though, we should note that
numerical simulations of the chromosphere and corona
sometimes have exhibited {\em shortfalls} of heating at the TR.
For example, both \citet{Wa94} and \citet{Vd10} found that producing
accurate models of the high-speed solar wind seemed to require
both an extended source of heat---arising presumably from
Alfv\'{e}nic turbulence---and an additional (ad~hoc) localized
source of heat near the coronal base.
\citet{Lg08} also found that adding an extra source of basal heating
was required to understand observations of the on-disk counterparts
of Type~II spicules.
\citet{SC16} were only able to model the thermodynamic properties of
coronal ``down-loops'' (i.e., loops in which the temperature decreases
toward the apex) by adding a heat source that behaves like an
impulsive source of compressive MHD wave energy at the TR.

\subsection{Inputs Describing the Magnetic Carpet}
\label{sec:TRopen:carpet}

We make use of an existing set of Monte Carlo simulations of
photospheric magnetic-flux transport and potential-field
extrapolation into the low corona.
\citet{CvB10} developed these models initially to determine the
rates at which closed field lines open up, and to estimate the
rate of energy release by interchange reconnection in such events.
\citet{Cr18} extended these models to predict the properties of
reconnection-driven MHD waves.
We continue to use these same models of the mixed-polarity
``magnetic carpet,'' but we also note that more recent high-resolution
data may point to the need for updating some parameters of these models
\citep[e.g., magnetic recycling times may be shorter; see][]{Wi13,Wa20}.

\begin{figure}[!t]
\vspace*{0.05in}
\epsscale{1.15}
\plotone{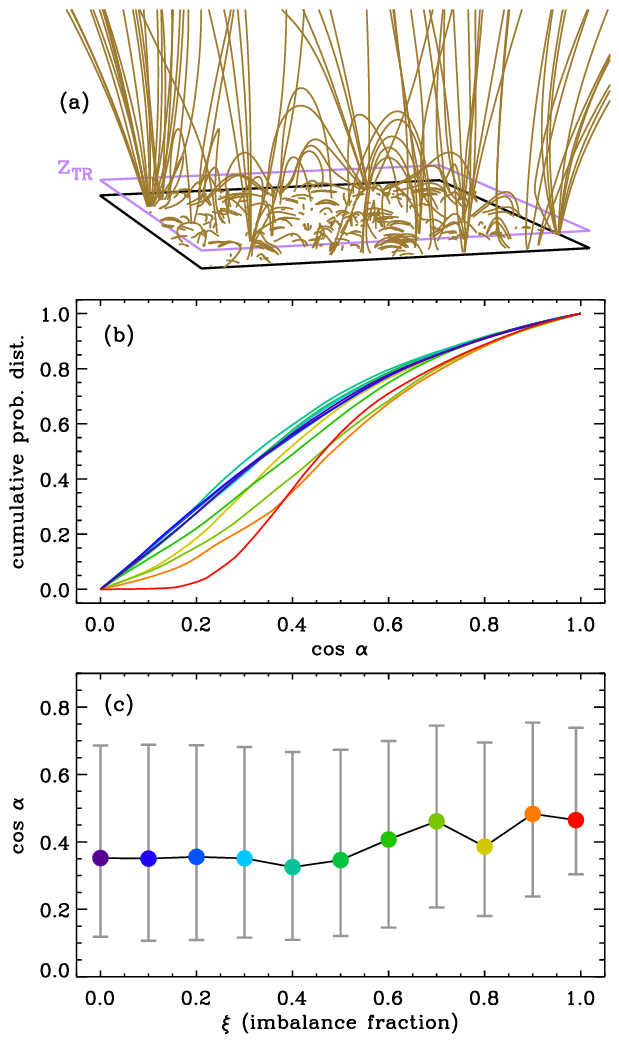}
\caption{(a) Perspective rendering of field lines for an example
timestep in the $\xi = 0.4$ model.
The height of the TR is illustrated by the purple outline.
(b) Cumulative probability distributions of $\cos\alpha$ for
each of the 11 models.
(c) Median (50\% percentile) and $\pm 1$ standard deviation limits
(16\% and 84\% percentiles) for the distributions of $\cos\alpha$
for each value of $\xi$.
Curve colors in (b) agree with symbol colors in (c).
\label{fig04}}
\end{figure}

Figure \ref{fig04}(a) shows a typical snapshot of the field-line
geometry for one of these Monte Carlo simulations.
The photospheric domain was assumed to be a square box 200~Mm
on a side, and it was filled with point-like magnetic elements
with positive and negative fluxes given by integer multiples of
$10^{17}$~Mx.
These models were evolved over several months of simulation time
and have reached a dynamical steady state.
The main free parameter is the large-scale magnetic flux imbalance
fraction $\xi$, which is defined as the ratio of the net flux density
($|B_{+}| - |B_{-}|$) to the absolute unsigned flux density
($|B_{+}| + |B_{-}|$).
Note that models with $\xi \lesssim 0.3$ are meant to simulate
``balanced'' patches of quiet Sun, whereas models with $\xi \gtrsim 0.7$
are meant to simulate more unipolar coronal-hole regions.

We compared the properties of 11 distinct Monte Carlo models with flux
imbalance ratios $\xi = 0,$ 0.1, 0.2, 0.3, 0.4, 0.5, 0.6, 0.7,
0.8, 0.9, and 0.99.
For each, we extracted a large ensemble of 500,000 random samples for
the field-line inclinations (i.e., values of $\alpha$
and $\beta$) at a typical transition-region height 
$z_{\rm TR} = 2.5$~Mm above the photosphere.
The distributions of $\beta$ azimuthal angles always appeared to be
reasonably uniform.
Figure~\ref{fig04}(b) shows the cumulative distributions of values
of $\cos\alpha$ for all 11 models.
They appear somewhat isotropic, but slightly biased more towards
horizontal orientations.
Note that a truly isotropic distribution of field-line vectors would 
correspond to a straight line in Figure~\ref{fig04}(b) ($y=x$) and thus
would have a median value of $\cos\alpha = 0.5$, or a median angle
$\alpha = 60^{\circ}$.

Figure \ref{fig04}(c) shows that the actual median values of $\cos\alpha$
take on values between 0.325 and 0.483 (i.e., corresponding to values
of $\alpha$ between $71^{\circ}$ and $61^{\circ}$, respectively),
with large standard deviations.
The most imbalanced ($\xi \geq 0.7$) configurations tend to have
more vertically oriented fields than those with balanced fluxes,
but the differences are subtle.
However, the departures from vertical fields are important to consider,
since an Alfv\'{e}n wave incident at an interface with
$\cos\alpha = 1$ (i.e., $\alpha = 0$) would produce no fast or slow
magnetosonic waves in either reflection or transmission.

Each model run in a given Monte Carlo ensemble is constructed with
randomly selected values of $\alpha$ and $\beta$, and these fully
determine the direction of ${\bf B}_0$ at the interface.
The $\beta$ angles are sampled from a uniform distribution
(between 0 and 360$^{\circ}$), and the $\alpha$ angles are sampled
from one of the distributions shown in Figure~\ref{fig04}(b).
The magnitude $B_0$ could also be sampled from observed or
simulated distributions, but for simplicity we chose to select a
single representative value to apply for each ensemble.
Then, later we compare the results to other ensembles with different
values of $B_0$.
By varying the magnetic-field strength over several orders of magnitude
between 0.1 and 1000 G, we can probe the behavior of TR-adjacent waves
for the full range of quiet to active regions in the solar atmosphere.

For the purposes of the \citet{St71} interface model, we assume the
gas pressure is constant across the TR, with
\begin{equation}
  \frac{\rho_1}{\rho_2} \, = \, \frac{T_2}{T_1} \, = \, 30
\end{equation}
and the chromospheric values (on the side from which the incident
wave comes) are $\rho_1 = 3 \times 10^{-14}$ g~cm$^{-3}$ and
$T_1 = 10^4$~K.
Note that typical quoted temperatures for the chromosphere ($10^4$~K) and
corona ($10^6$~K) would presume the existence of a larger-magnitude jump
of order $T_2/T_1 \approx 100$.
However, the above value of 30 corresponds to only the lowest and
steepest part of the TR that subtends no more than a few hundred km
\citep[see, e.g.,][]{AL08}.
Because the temperature in the uppermost part of the TR rises more gradually,
this region should not be counted as part of the sharp interface.

The final parameters to set, for each run in a Monte Carlo ensemble,
are the frequency $\omega$ and wavenumber ${\bf k}$ of the incident
Alfv\'{e}n wave.
Because the interface is assumed to be infinitely thin, the actual
value of the wave frequency does not affect the calculation of
reflection, transmission, and mode-coupling.
To keep the values of all parameters realistic, we chose a
representative value of $\omega = 0.02$ rad~s$^{-1}$, corresponding
to incident Alfv\'{e}n-wave periods of about 5 minutes
\citep[see, e.g.,][]{To07}.
The wavenumber is specified by the dispersion relations and by
$\theta_k$, which we initially sample randomly between
0 and $180^{\circ}$.
However, we accept or reject each trial value depending on the
resulting value of $\Theta_{kB}$ for the incident Alfv\'{e}n wave.
We explore three distinct models of turbulent anisotropy that we
define below:
\begin{enumerate}
\item
Model I (isotropic): as long as $F_z > 0$ for the incident wave,
all values of $\theta_k$ are retained.
\item
Model S (slab): only values with $F_z > 0$ and
$|\cos\Theta_{kB}| \geq 0.9848$ (i.e., quasi-parallel orientations
of $\bf k$ within $10^{\circ}$ of ${\bf B}_0$) are retained.
\item
Model T (transverse): only values with $F_z > 0$ and
$|\cos\Theta_{kB}| \leq 0.1736$ (i.e., $\bf k$ within $10^{\circ}$
of the plane perpendicular to ${\bf B}_0$) are retained.
\end{enumerate}
Note that despite many possible sampled values being rejected in
models S and T, we keep resampling until the desired total number of
trials is achieved.

In summary, each run in a given Monte Carlo ensemble involves sampling
three variables from random distributions
($\alpha$, $\beta$, and $\theta_k$) and keeping others fixed
(e.g., $\xi$, $B_0$, $\rho_2/\rho_1$, $T_2/T_1$).
Typically, each ensemble was constructed with 30,000 trials, using
the Interactive Data Language's standard Mersenne Twister algorithm
for pseudorandom number generation \citep{MN98}.

\subsection{Mode Coupling Results}
\label{sec:TRopen:fluxes}

We begin by discussing the computed distributions of reflection and
transmission coefficients (for all three types of MHD waves) for a
typical Monte Carlo ensemble run with
$B_0 = 10$~G, the distribution of $\cos\alpha$ angles taken from
the $\xi = 0.7$ model, and an isotropic distribution
of $\theta_k$ angles (i.e., Model~I).
The adopted value of the magnetic field corresponds to a plasma beta
ratio (gas pressure to magnetic pressure) of about 0.01.

\begin{figure}[!t]
\vspace*{0.05in}
\epsscale{1.15}
\plotone{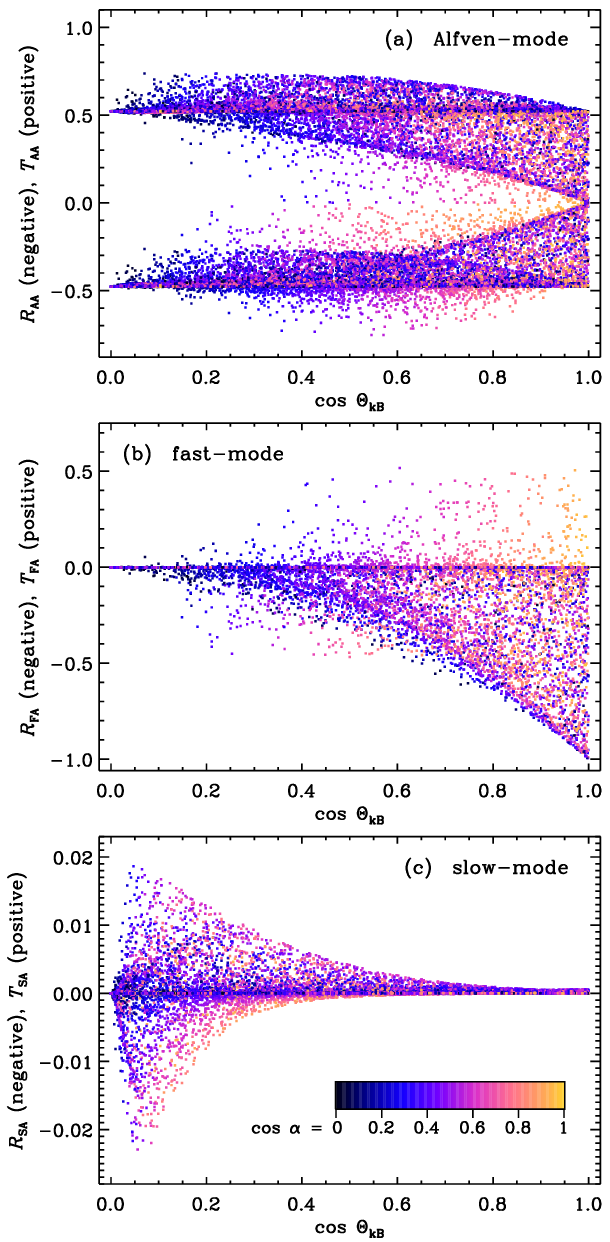}
\caption{(a) Transmission (positive) and reflection (negative)
coefficients for Alfv\'{e}n waves produced at the TR for a
Model-I Monte Carlo ensemble with $B_0 = 10$~G and $\xi = 0.7$.
(b) Same as panel~(a), but for fast-mode MHD waves.
(c) Same as panel~(a), but for slow-mode MHD waves.
Symbol color denotes $\cos\alpha$, as shown in the color-bar.
\label{fig05}}
\end{figure}

Figure~\ref{fig05}(a) shows the computed distributions of
${\cal R}_{\rm AA}$ and ${\cal T}_{\rm AA}$ values, displaying
reflection coefficients as negative values for the sake of clarity.
In a majority of cases, the computed values are quite close to those
predicted by Equation~(\ref{eq:RAA1}), i.e.,
${\cal R}_{\rm AA} = 0.478$ and ${\cal T}_{\rm AA} = 0.522$,
which were computed assuming $\beta = 0$.
In fact, 69\% of the 30,000 trials have a value of ${\cal R}_{\rm AA}$
that falls within 10\% of its corresponding analytic value, and
67\% of the trials have a value of ${\cal T}_{\rm AA}$
that falls within 10\% of its analytic value.
The remaining trials---amounting to roughly one-third of the
total number---exhibit values that often indicate substantial
coupling with the other modes.

Figures \ref{fig05}(b) and \ref{fig05}(c) show how the reflection
and transmission coefficients of fast and slow magnetosonic waves
behave as a function of the incident Alfv\'{e}n wave's $\Theta_{kB}$
angle and the magnetic field's $\alpha$ angle.
In the majority of cases, these four coefficients have quite small values,
pointing to the relative weakness of mode coupling.
We computed the total flux fractions going into non-Alfv\'{e}nic modes,
i.e.,
\begin{equation}
  f \, = \, {\cal R}_{\rm FA} + {\cal T}_{\rm FA} +
            {\cal R}_{\rm SA} + {\cal T}_{\rm SA}
\end{equation}
and, although $f$ sometimes grows to order unity, 
the median value of $f$ (for the full distribution of 30,000 trials)
was only 0.00081.
Note also that incident waves propagating parallel to the magnetic
field ($\cos\Theta_{kB} \approx 1$) tended to couple the most strongly
to the fast mode, and incident waves propagating transversely to
the field ($\cos\Theta_{kB} \approx 0$) coupled strongly
to the slow mode.
This behavior makes sense in the context of a low-plasma-beta
environment, in which fast modes have similar phase speeds as
Alfv\'{e}n waves for nearly parallel propagation and
slow modes have similar phase speeds as Alfv\'{e}n waves for
nearly perpendicular propagation.
Matching phase speeds appears to be a necessary but not sufficient
condition for strong mode coupling \citep[see also][]{St71,Va90}.

When considering the expected energy fluxes of upward and downward MHD
waves that would be produced in the vicinity of the TR, we compute
{\em mean values} for the reflection and transmission coefficients
produced in each Monte Carlo ensemble.
This is meant to be a proxy for some degree of spatial and temporal
averaging over unresolved scales in the complex and rapidly evolving
magnetic carpet.
Thus, in the remainder of this section, we report only such mean values.

\begin{figure}[!t]
\vspace*{0.05in}
\epsscale{1.15}
\plotone{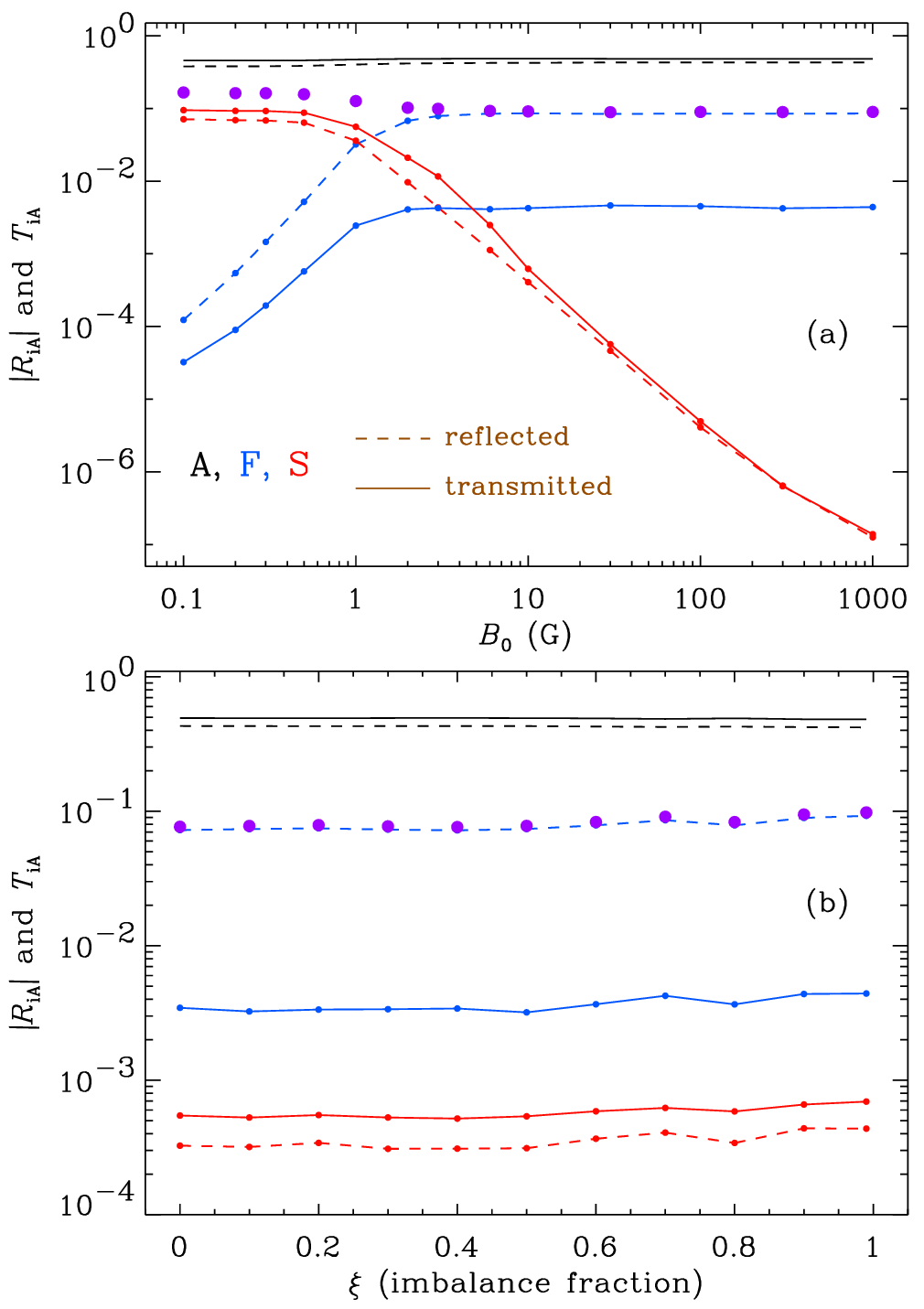}
\caption{Ensemble-averaged reflection coefficients (dashed curves)
and transmission coefficients (solid curves),
with Alfv\'{e}n, fast, and slow wave properties shown in black, blue,
and red, respectively.
(a) Dependence on $B_0$ for $\xi = 0.7$ and Model~I.
(b) Dependence on $\xi$ for $B_0 = 10$~G and Model~I.
Values of $f$ (i.e., totals for all four non-Alfv\'{e}nic components)
are shown with purple symbols.
\label{fig06}}
\end{figure}

Figure \ref{fig06}(a) shows how the mode-coupling coefficients behave
as a function of $B_0$, for fixed choices of
$\xi = 0.7$ and isotropic incident wavenumbers (Model~I).
As in the example above, for the strongest magnetic-field cases
(i.e., the lowest plasma beta ratios) there is a much more efficient
energy transfer from Alfv\'{e}n to fast-mode waves than to slow-mode waves.
This trend reverses itself for the weakest fields (i.e., highest
plasma beta ratios), for which the coupling with slow-mode waves is
most efficient.
It is also interesting that there is always about an order of magnitude
higher amount of reflection for fast-mode waves than transmission
(i.e., ${\cal R}_{\rm FA}/{\cal T}_{\rm FA} \approx 10$),
but the slow-mode waves seem to be partitioned more equally between
the reflected and transmitted directions.

Figure \ref{fig06}(b) shows how the mean reflection and transmission
coefficients vary as a function of $\xi$, for fixed values of
$B_0 = 10$~G and isotropic incident wavenumbers (Model~I).
Varying $\xi$ changes the probability distribution of
$\cos\alpha$ values from which the Monte Carlo algorithm draws
(see Figure~\ref{fig04}(b)).
It is somewhat surprising how little the mean coefficients change
as one goes from balanced quiet-Sun patches ($\xi \approx 0$)
to highly imbalanced coronal-hole regions ($\xi \approx 1$).
It is possible, of course, that there exist other differences between
these regions---such as differences in the mean height of the TR
or different mean values of $\rho$ or $T$---that would produce
even larger variations in the reflection and transmission coefficients,
had they been included.

\begin{figure}[!t]
\vspace*{0.05in}
\epsscale{1.15}
\plotone{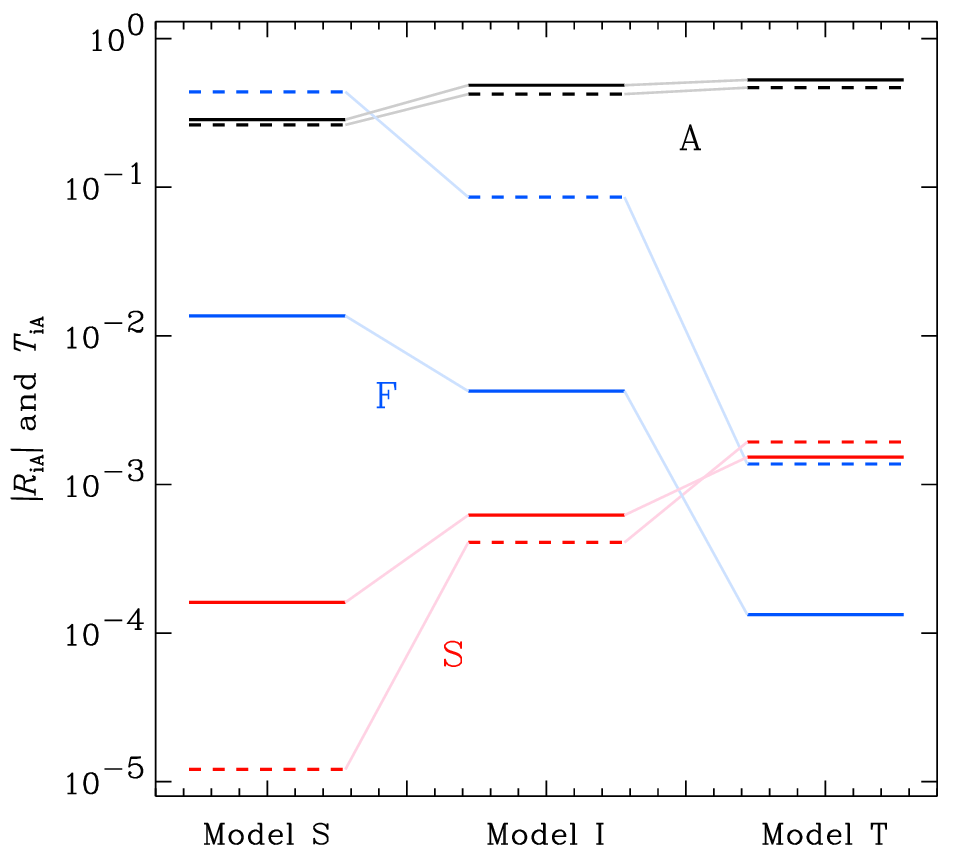}
\caption{Ensemble-averaged reflection and transmission coefficients,
with line styles and colors the same as in Figure~\ref{fig06}.
In order to illustrate the dependences on wavenumber anisotropy,
we use dotted lines to connect identical parameters.
\label{fig07}}
\end{figure}

Lastly, for the standard set of fixed parameters
($B_0 = 10$~G and $\xi = 0.7$), we vary the wavenumber anisotropy
model and show the results in Figure~\ref{fig07}.
It makes sense that Model~S produces the most intense coupling
between Alfv\'{e}n and fast-mode waves, since these two modes are
the most similar to one another for parallel propagation
(i.e., $\Theta_{kB} \rightarrow 0$).
For the other extreme case of perpendicular propagation (Model~T),
the reflected fast-mode and slow-mode waves have roughly equal fluxes,
and the transmitted slow-mode waves have about an order of magnitude
larger flux than the transmitted fast-mode waves.

\subsection{Collisional Heating Rates}
\label{sec:TRopen:heating}

In the solar atmosphere, all three types of MHD waves undergo dissipation
due to particle-particle collisions \citep[see, e.g.,][]{A47,O61}.
The specific processes at work include thermal conductivity,
viscosity, electrical resistivity (i.e., Ohmic or Joule dissipation),
and a range of ion–neutral couplings.
Both slow-mode and fast-mode magnetosonic waves tend to damp out
more rapidly than do Alfv\'{e}n waves, which explains the popularity of
the latter as proposed sources of energy deposition in the extended
corona.
However, our goal in this section is to simulate the dissipation of
the newly produced slow-mode and fast-mode waves.
This damping appears to produce a localized ``bump'' of heating in
the vicinity of the TR that may be an important contributor to the
overall coronal heating problem.

\begin{figure}[!t]
\vspace*{0.05in}
\epsscale{1.15}
\plotone{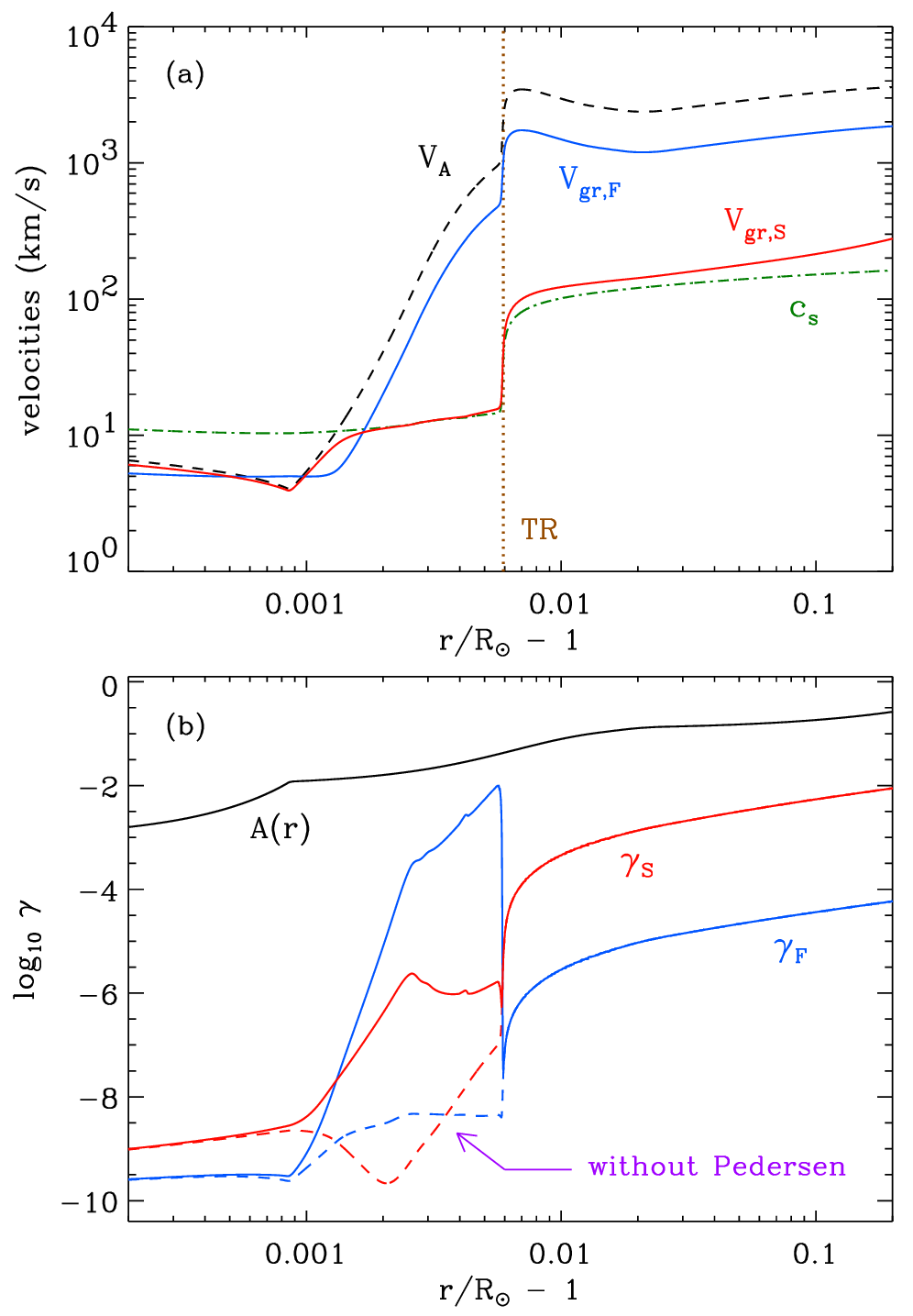}
\caption{(a) Radial dependence of characteristic velocities such
as the sound speed $c_s$ (green dot-dashed curve),
Alfv\'{e}n speed $V_{\rm A}$ (black dashed curve),
and angle-averaged group velocities for fast-mode (blue solid curve)
and slow-mode (red solid curve) waves.
(b) Radial dependence of flux-tube area $A$ (black solid curve)
and collisional damping rates for fast-mode (blue) and
slow-mode (red) waves,
with both total values computed with Pedersen conductivity
(solid curves) as well as the fully-ionized Braginskii limit
(dashed curves).
\label{fig08}}
\end{figure}

We compute the radial variation of wave energy density $U_i$,
where the subscript $i$ is either F or S for fast-mode or slow-mode
waves, respectively.
We use the time-steady wave-action conservation equation described by
\citet{CvB12}, and for simplicity we provide it here only the limit of
a negligibly small solar-wind speed (which is valid near the TR).
Thus,
\begin{equation}
  \frac{1}{A} \frac{d}{dr} \Big( A V_{{\rm gr},i} U_i \Big)
  \, = \, -2 \gamma_i U_i \,\,\, ,
  \label{eq:action}
\end{equation}
where $A$ is the cross-sectional area of a presumed vertically oriented
magnetic flux tube (i.e., $A \propto 1/B_0$),
$V_{{\rm gr},i}$ is the group velocity of mode $i$, and
$\gamma_i$ is its total linear damping rate.
We also define the total heating rate due to damping as
$Q_i = 2 \gamma_i U_i$.
Since the above equation will be applied most often to the isotropic
distribution of wavenumbers (Model~I), we use the angle-averaging
procedure described in Section~2.3 of \citet{CvB12} to compute
representative values of $V_{{\rm gr},i}$.
Figure~\ref{fig08}(a) shows how some of these quantities depend on
height (measured as distance above the solar photosphere in units of
the solar radius, $R_{\odot}$) in the vicinity of the TR.
To set the background properties of the atmosphere, we chose the
coronal-hole ZEPHYR model of \citet{CvB07} that used a photospheric
acoustic flux of 1 kW~m$^{-2}$.

For a fully ionized plasma, the linear damping rates due to viscosity,
thermal conduction, and Ohmic resistivity were discussed in detail
by, e.g., \citet{B65} and \citet{Wh97}.
We use the specific forms given in Appendix~B of \citet{CvB12}, which
include modification terms for weak collisions.
These were derived in order to avoid the ``molasses limit'' of the
classical expressions; i.e., an unphysical divergence of the transport
coefficients when the mean time between collisions becomes infinite
\citep[see also][]{Wi95}.
However, for a plasma that is only partially ionized, the presence of
neutral atoms produces additional resistivity via Pedersen currents
\citep{Goo04,Goo11,Kaz06}.
This effect must be included when modeling the chromosphere and TR.
Thus, the modified version of the total damping rate is given by
the sum of viscous, Ohmic, and conductive terms,
\begin{equation}
  \gamma_{i} \, = \,
  \gamma_{{\rm vis,}i} +
  \gamma_{{\rm ohm,}i} (1 + \Gamma) +
  \gamma_{{\rm con,}i} \,\,\, .
\end{equation}
Note that the conductive damping term $\gamma_{{\rm con,}i}$
is identically zero for the purely incompressible Alfv\'{e}n mode, but
it is a non-negligible contributor to the total damping rate for the
fast and slow mode waves.
The Pedersen correction term $\Gamma$, due to the presence of
neutrals, is discussed in more detail by \citet{Kaz06} and is given by
\begin{equation}
  \Gamma \, = \, \left( \frac{\rho_{\rm HI}}{\rho_{\rm H}} \right)^2
  \, \frac{\Omega_e \Omega_p}{\nu_e^{\ast} \nu_p^{\ast}}
\end{equation}
where $\Omega_p$ and $\Omega_e$ are the cyclotron frequencies of
electrons and protons, and $\nu_e^{\ast}$ and $\nu_p^{\ast}$
are the reduced collision frequencies defined by \citet{Goo04}.
We specify the neutral hydrogen density fraction
$\rho_{\rm HI}/\rho_{\rm H}$ with the same temperature-dependent
tabulation that was used by \citet{CvB07}, which has a similar
behavior as more realistic time-dependent models
\citep[e.g.,][]{Ca16,Pz22}.

Figure~\ref{fig08}(b) shows the resulting height dependence of
$\gamma_{\rm F}$ and $\gamma_{\rm S}$, both with and without the
inclusion of the Pedersen term.
We find, as expected, that $\Gamma \ll 1$ in the corona, but there is
a strong enhancement in both damping rates (due to $\Gamma \gg 1$)
in the upper chromosphere.
Note that the damping rates also depend on the square of the
wavenumber, so there is an associated dependence on the
wave frequency.
As above, we chose a typical value of
$\omega = 0.02$ rad~s$^{-1}$, corresponding to a period of 5 minutes,
and we discuss the dependence on $\omega$ further below.

In order to solve Equation~(\ref{eq:action}) both above and below the
TR, there are several additional quantities to specify.
We are assuming that mode coupling produces the ``new'' fast-mode and
slow-mode waves precisely at the TR interface.
Thus, we must use the transmission coefficients to
determine the boundary conditions on $U_i$ for integrating upwards,
and we must use the reflection coefficients to determine the boundary
conditions on $U_i$ for integrating downwards.
The interface is defined as the location at which the time-steady
model temperature $T(r)$ is equal to
$(T_1 T_2)^{1/2} = 5.48 \times 10^{4}$~K, and for this model it is
found at a height $r_{\rm TR}$ of 0.0059~$R_{\odot}$ (i.e., 4,100~km)
above the photospheric base.
The incident Alfv\'{e}n wave is assumed to have a known vertical
energy flux $F_{z,{\rm A}}$ at the interface, and various observational
constraints have provided values of this quantity in the range of
roughly 3--6~kW~m$^{-2}$ \citep{CvB05,CvB12}.
We choose the lower bound of that range (3~kW~m$^{-2}$) so as not to
overestimate the heating-rate predictions of this paper.
Using this value, the boundary conditions for the fluxes
of transmitted and reflected magnetosonic waves are given as
either $F_{z,i} = F_{z,{\rm A}} {\cal T}_{i{\rm A}}$ (for upward waves)
or $F_{z,i} = F_{z,{\rm A}} {\cal R}_{i{\rm A}}$ (for downward waves),
and the energy densities are given as
\begin{equation}
  U_i \, = \, F_{z,i} / V_{{\rm gr},i}
\end{equation}
which we solve four times (for fast and slow modes; in the upward
and downward directions) at the interface.

\subsection{Localized Heating Results}
\label{sec:TRopen:results}

Equation~(\ref{eq:action}) was solved numerically by integrating
both upwards and downwards from the TR.
Because of the sharp gradients in this region, we found that 
straightforward finite-difference integration steps were often unstable,
even with very finely spaced grids in radial distance.
Thus, we took advantage of the form of the differential equation and
noted that it can be written as
\begin{equation}
  \frac{dy}{y} \, = \, -\frac{dr}{H}
\end{equation}
where $y = A V_{{\rm gr},i} U_i$ and $H = V_{{\rm gr},i} / 2 \gamma_i$.
This equation has a known solution for $H \approx$~constant, so a
reasonably robust finite-differencing scheme, going from step $(n)$
to step $(n+1)$, was found to be
\begin{equation}
  y^{(n+1)} \, = \, y^{(n)} \exp \left(
  -\frac{r^{(n+1)}-r^{(n)}}{H^{(n)}} \right)  \,\,\, .
\end{equation}
Once $y(r)$ is determined, we compute $U_i(r)$ for each mode,
as well as the corresponding rms velocity amplitudes
$\delta v_i = (U_i/\rho)^{1/2}$.
This expression makes use of the fact that linear MHD waves carry
exactly half of their total fluctuation energy in the form of kinetic
energy \cite[see, e.g.,][]{GP04}.

\begin{figure}[!t]
\vspace*{0.05in}
\epsscale{1.15}
\plotone{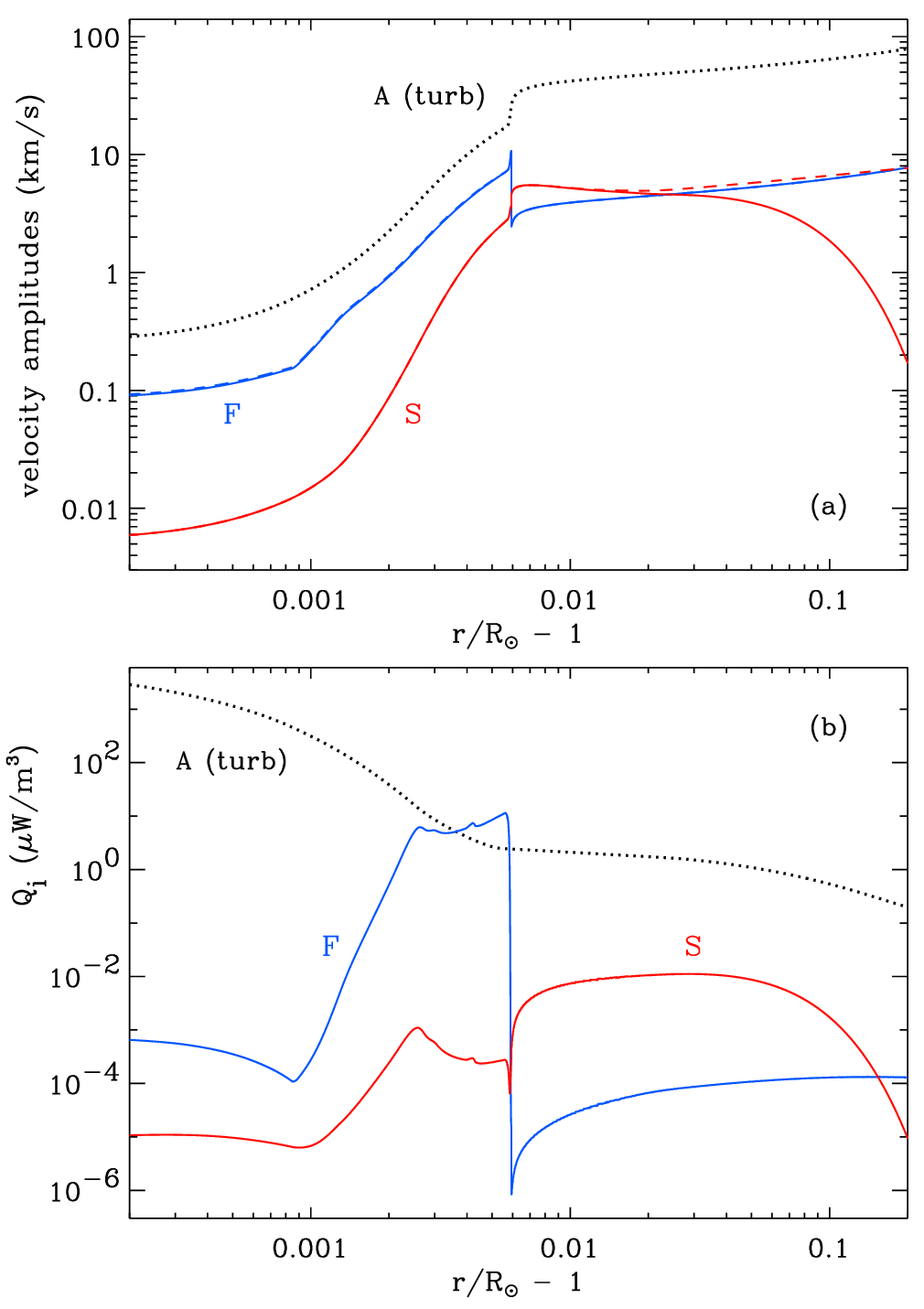}
\caption{(a) Radial dependence of wave velocity amplitudes
for fast-mode waves (blue curves), slow-mode waves (red curves), and
Alfv\'{e}nic turbulence from the time-steady model (black dotted curve).
For magnetosonic modes, both undamped (dashed) and damped (solid)
amplitudes are shown.
(b) Radial dependence of volumetric heating rates due to
fast-mode waves (blue curves), slow-mode waves (red curves),
and Alfv\'{e}nic turbulence (black dotted curve).
\label{fig09}}
\end{figure}

Figure~\ref{fig09}(a) shows the computed radial dependences of wave
velocity amplitudes for the typical case shown in Figure~\ref{fig05};
i.e., $B_0 = 10$~G, $\xi = 0.7$, and an isotropic distribution
of $\theta_k$ angles (Model~I).
We show both the full solutions for slow-mode and fast-mode
amplitudes and also those computed without any damping (i.e.,
assuming $\gamma_i = 0$).
Note that the upward slow-mode waves become very strongly damped in
the corona, whereas the downward slow-mode waves experience almost no
damping (with the damped amplitudes only decreasing to 0.997 times
the undamped amplitudes).
The downward fast-mode waves have the largest amplitudes of the four
magnetosonic waves generated at the TR, and they are also damped rather
weakly (only down to 0.98 times the corresponding undamped amplitudes).
Also shown are the transverse velocity amplitudes of incompressible
Alfv\'{e}nic waves that are computed self-consistently by the
time-steady ZEPHYR model of \citet{CvB07}.
These rms amplitudes account for fluctuations in
both dimensions transverse to the vertical magnetic field.
Thus, the values of $\sim$50~km~s$^{-1}$ in the low corona correspond
to projected nonthermal line widths of order
$50 / \sqrt{2} \approx 35$~km~s$^{-1}$, which agrees with
off-limb measurements \citep[e.g.,][]{Bj98}.

The corresponding heating rates $Q_i$, expressed in units of
$\mu$W~m$^{-3}$, are shown in Figure~\ref{fig09}(b).
Despite the relatively weak damping for the downward fast-mode waves,
they contribute the most to the heating due to their large amplitudes.
In fact, it is this component that is most responsible for the
compact ``bump'' of extra heating due to newly produced waves at
the TR interface.
For the model parameters shown here, the peak of the bump exceeds the
background heating---due Alfv\'{e}nic turbulence in the ZEPHYR model---by
about a factor of five.
This extra heating has been computed without concern for how it would
self-consistently alter the presume background plasma state.
Given that the extra heating occurs {\em below} the modeled TR,
we anticipate that including it in the time-steady energy balance
would {\em lower} the height of the TR.
This may help explain why prior generations of the ZEPHYR code
\citep{CvB07,CvB13} found the TR to occur at heights substantially
above those seen in observationally guided models
\citep[e.g.,][]{FAL90,AL08}.

\begin{figure}[!t]
\vspace*{0.05in}
\epsscale{1.15}
\plotone{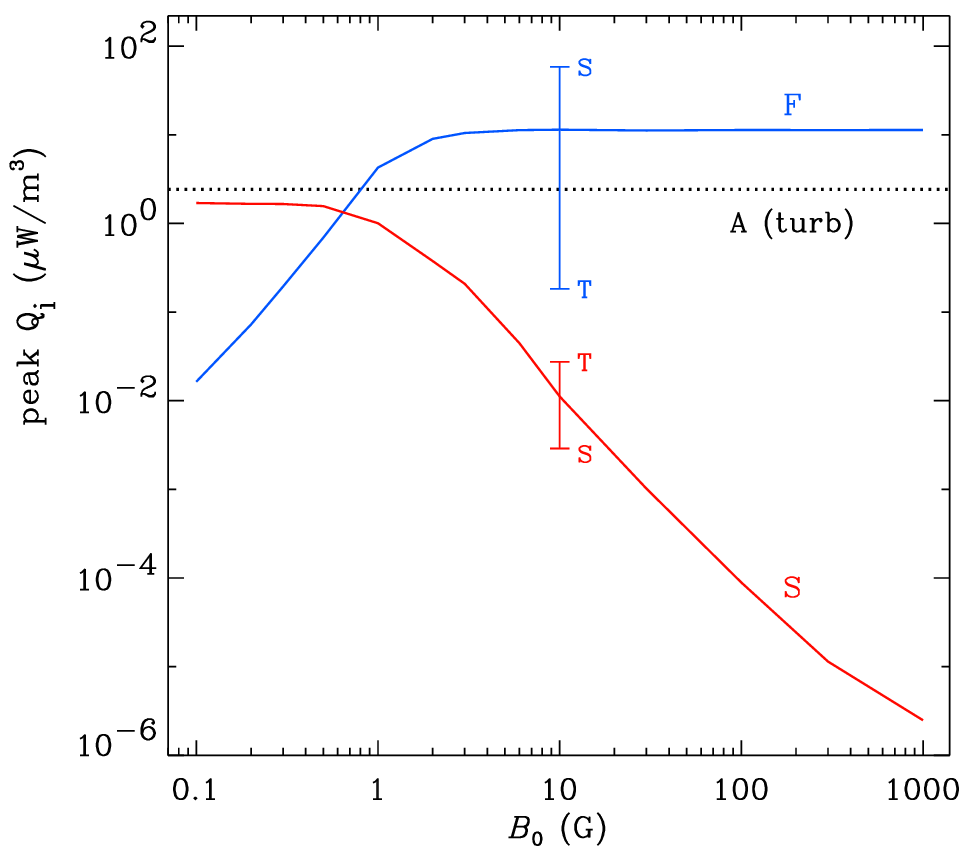}
\caption{Peak heating rates due to the damping of
fast-mode waves (blue curve) and slow-mode waves (red curve),
produced at the TR, as a function of $B_0$ for $\xi=0.7$ and Model~I.
Vertical struts at $B_0 = 10$~G show how the peak heating rate changes 
when other wavenumber anisotropy models are applied (Models S or T).
The typical heating rate due to incompressible Alfv\'{e}nic turbulence
at the TR is also shown (black dotted curve).
\label{fig10}}
\end{figure}

Figure~\ref{fig10} shows how the peak value of $Q_i$ depends on $B_0$.
Here, we continue to assume $\xi=0.7$ and the wavenumber
anisotropy properties of Model~I, as in Figure \ref{fig06}(a).
Heating due to newly produced magnetosonic waves at the TR remains
dominated by downward-propagating fast-mode waves for field strengths
exceeding just a few Gauss.
In regions with even weaker fields, the peak heating rate is driven
mainly by the damping of both upward and downward slow-mode waves.
The vertical struts show how the model with $B_0 = 10$~G exhibits
either stronger or weaker heating when the wavenumber anisotropy
conditions are varied to Models S and T (see also Figure~\ref{fig07}).
Most notably, the assumption that most of the upflowing Alfv\'{e}n waves
have wavenumbers nearly parallel to the magnetic field (i.e., Model~S)
seems to produce a peak heating rate at the TR that is about 25 times
larger than the expected level of background heating from incompressible
turbulence.
(In a completely self-consistent model, the heating due to
Alfv\'{e}nic turbulence would vary with $B_0$ as well, but here we
treat this as an aspect of the ``background'' atmosphere.)

As mentioned above, the collisional heating rate $Q_i$ is proportional
to the square of the assumed wave frequency.
When varying $\omega$ up and down by several orders of magnitude
from the baseline value of 0.02 rad~s$^{-1}$, we found that the
radial profile of $Q_i(r)$ simply shifted up and down by the expected
amount and did not display any noticeable changes in shape.
However, our assumption of a single monochromatic frequency is not
likely to be realistic.
The Sun appears to exhibit continuous power spectra of wave energy
flux as a function of frequency.
If this flux spectrum is a power-law with $P \propto \omega^{-n}$,
then the spectrum of associated heating rates would go as
$Q \propto \omega^{2-n}$, and its integral over frequency would
behave as $\omega^{3-n}$.
Thus, if $n > 3$ the total heating would be dominated by the
low-frequency end of the spectrum, and if $n < 3$ it would be
dominated by the high-frequency end of the spectrum.
Unfortunately, observations of both chromospheric and coronal waves
do not show a consensus, with reported values of the power-law slopes
varying between 1 and 4 for different regions and diagnostics
\citep[see, e.g.,][]{Re08,TM09,Mo21,Mo23}.

For the models with $B_0 \gtrsim 3$~G discussed above (i.e., the
cases where the heating is dominated by downward-propagating
fast-mode waves), the following scaling relation provides a good
way to estimate the peak heating rate at the TR:
\begin{equation}
  \mbox{max}(Q_{\rm F}) \, \approx \,
  \frac{F_{z,{\rm A}} \langle {\cal R}_{\rm FA} \rangle}
  {\lambda_{\rm eff}}
  \left( \frac{\mbox{5 min}}{\cal P} \right)^2 \,\,\, ,
\end{equation}
where ${\cal P}$ is the wave period,
$\langle {\cal R}_{\rm FA} \rangle$ is the ensemble-averaged
reflection coefficient for fast-mode waves, and we define an effective
length scale $\lambda_{\rm eff} = 0.034 \, R_{\odot}$ to set the
overall normalization.
Other models of the background plasma and ionization state will
exhibit different values of $\lambda_{\rm eff}$, but the
other scalings given above should remain valid.

\section{Discussion and Conclusions}
\label{sec:conc}

In this paper we have explored a few of the possible ramifications
of Alfv\'{e}n waves in the solar atmosphere encountering sharp 
discontinuities in density.
In Section~\ref{sec:alf}, we found that linear reflection from
multiple interfaces may help explain simulated enhancements in
sunward-propagating Alfv\'{e}nic turbulence in the presence of
density fluctuations.
Over the past decade, it has been realized that purely incompressible
reflection-driven turbulence may not be sufficient to heat the
extended corona and accelerate the solar wind on its own
\citep[e.g.,][]{vB16}.
Without compressible effects, the inward-to-outward Elsasser
amplitude ratio $Z_{-}/Z_{+}$ may not exceed values of order 0.01.
However, there is a great deal of observational evidence for
density fluctuations with magnitudes $\delta \rho / \rho_0$
exceeding 0.1 in the corona and solar wind
\citep{Is98,Sp02,Miy14,Hh18,Wx19,Mh21,Cu23}.
The associated reflection due to these fluctuations may enhance
the total amount of Alfv\'{e}nic reflection to values of order
$Z_{-}/Z_{+} \gtrsim 0.1$ (see Figure~\ref{fig03}).
Those more intense levels of sunward-propagating waves are much
more likely to provide sufficient levels of turbulent heating for
the corona and solar wind.

In Section~\ref{sec:TRopen}, we found that taking account of a
realistic distribution of magnetic field directions (i.e., the highly
structured ``magnetic carpet'') may allow upward-propagating Alfv\'{e}n
waves to convert a non-negligible fraction of their energy into both
fast-mode and slow-mode MHD waves at the sharp TR.
These newly produced waves dissipate efficiently in the vicinity of
the TR and produce a narrow peak of heating that, in some cases, may
exceed the dominant rate of turbulent heating by an order of magnitude.
These little bumps of extra heating are not likely to be responsible
for the majority of chromospheric or coronal heating,
but their existence should not be ignored.
In fact, this heretofore unknown source of heating may be an explanation
for why some models have not been able to produce realistic coronal
temperature distributions without adding sharp (but essentially ad-hoc)
sources of heat at this location \citep[e.g.,][]{Wa94,Lg08,Vd10,SC16}.

It is important to note that our prediction of a narrow bump of additional
heating at the TR is not based on a fully self-consistent simulation.
A more comprehensive understanding of this effect will require taking into
account some additional processes and complexities, such as the following.
\begin{enumerate}
\item
More realistic geometrical parameters for the system should be adopted.
This includes the use of self-consistent distributions of magnetic-field
directions and magnitudes---say, from simulations like Bifrost
\citep{Ca16,Mo22}---and also a naturally ``corrugated'' TR interface
between the chromosphere and corona \citep{Fe79,Pe13}.
It may also be useful to separate the closed and open magnetic
field lines to study differences in wave propagation between these
two regions.
\item
If the Alfv\'{e}n waves coming from below are already turbulent by the
time they reach the TR (and many simulations show they are; see
\citeauthor{vB11}\  \citeyear{vB11}), then their wavenumber angles
$\theta_k$ should be sampled from a probability distribution that
is consistent with the behavior of anisotropic MHD turbulent cascade.
\item
The physics of wave dissipation likely needs to be treated in a more
comprehensive manner.
Besides Pedersen currents, several other channels of ion-neutral
interaction have been proposed to be important in the chromosphere
\citep[e.g.,][]{Kh18,MS20}.
Also, the collisional transport coefficients may also be higher if the
electrons, ions, and neutral atoms have non-Maxwellian velocity
distributions \citep{Hs21}.
Lastly, our assumption of a simple temperature-dependent
ionization-recombination equilibrium, which includes only hydrogen,
should be replaced by more self-consistent physics (again,
see codes like Bifrost or MuRAM; \citeauthor{Ca16} \citeyear{Ca16};
\citeauthor{Pz22} \citeyear{Pz22}).
\item
This work assumed that the only waves incident to the TR interface
were Alfv\'{e}n waves coming up from below.
Of course, the real wave ecosystem of the solar atmosphere must also
involve magnetosonic waves coming up from below
\citep[e.g.,][]{O61,dW09,Js23} as well as coronal fluctuations
that propagate downwards (as we discussed in Section~\ref{sec:alf}).
The near-Sun solar wind is observed to contain all three types of
linear MHD waves, as well as other compressive fluctuations that are
not normal modes of a homogeneous plasma \citep[e.g.,][]{TM94,Zh21}.
We still do not know to what extent these waves come from
the solar surface or are generated in~situ.
\end{enumerate}

This paper has examined several solar applications of
Stein's (\citeyear{St71}) theory of MHD-wave interactions at a
sharp interface, and there are several others that could be
explored as well.
For example, closed loops in the low corona contain two TRs,
one at each footpoint, and their finite length is known to produce
resonances \citep{Io78,NV05} and ``sloshing oscillations'' \citep{Xi22}.
The seemingly infinite regress of multiple reflections and transmissions
from Section~\ref{sec:alf} may be extended to better understand the
behavior of trapped waves in these loops.
Further out in the open-field corona and solar wind, there exist
a wide range of field-aligned density striations \citep{Ra14,De16}.
\citet{MV22} suggested that interactions between Alfv\'{e}n waves
and these striations may result in the production of a $1/f$ power
spectrum of turbulent fluctuations, which is seen at low
frequencies in the solar wind.
Treating these striations as sharp interfaces may lead to new insights
about how they affect MHD waves in various environments.

\begin{acknowledgments}
The authors gratefully acknowledge
Bernie Vasquez and Munehito Shoda
for many valuable discussions.
The authors are also grateful to the anonymous referee for
many constructive suggestions that have improved this paper.
This work was supported by the National Aeronautics and Space
Administration (NASA) under grant 80NSSC20K1319,
and by the National Science Foundation (NSF) under grant 1613207.
The National Center for Atmospheric Research is a major facility
sponsored by the NSF under Cooperative Agreement No.\  1852977. 
M.E.M.\  was supported by the George Ellery Hale Graduate Student
Fellowship at the University of Colorado Boulder.
This research made extensive use of NASA's Astrophysics Data System (ADS).
\end{acknowledgments}

\newpage
\startlongtable
\begin{deluxetable}{ccccl}
\tablecaption{Turbulence Reflection Rates and Density Fluctuations
from Published MHD Simulations
\label{table01}}
\tablewidth{0pt}
\tablehead{
\colhead{$r/R_{\odot}$} &
\colhead{$\log (Z_{-}/Z_{+})$} &
\colhead{$\log (Z_{-}/Z_{+})_0$} &
\colhead{$\log (\delta \rho / \rho_0)$} &
\colhead{Source}
}
\startdata
    1.5 &  --1.2314 & --1.7175 & --1.0000 & vB16 \\
    2.4 &  --1.2098 & --1.6048 & --1.0000 & vB16 \\
    3.6 &  --0.9674 & --1.8117 & --1.0000 & vB16 \\
    5.5 &  --1.0028 & --1.9588 & --1.0000 & vB16 \\
    8.4 &  --0.9067 & --2.0498 & --1.0000 & vB16 \\
   13.0 &  --0.7228 & --2.0853 & --1.0000 & vB16 \\
   20.0 &  --0.7958 & --2.0816 & --1.0000 & vB16 \\
    2.0 &  --1.4367 & --1.6351 & --1.0458 & AT21, Fig.\  4(a,b)\\
    4.0 &  --0.9061 & --1.6842 & --0.6576 & AT21, Fig.\  4(a,b)\\
   10.0 &  --0.7341 & --1.7724 & --0.6198 & AT21, Fig.\  4(a,b)\\
   20.0 &  --0.7434 & --1.7570 & --0.6990 & AT21, Fig.\  4(a,b)\\
    2.0 &  --1.5607 & --1.6243 & --1.3979 & AT21, Fig.\  4(d,e)\\
    4.0 &  --0.8451 & --1.6690 & --0.6198 & AT21, Fig.\  4(d,e)\\
   10.0 &  --0.5699 & --1.7103 & --0.5229 & AT21, Fig.\  4(d,e)\\
   20.0 &  --0.8893 & --1.6920 & --0.9208 & AT21, Fig.\  4(d,e)\\
    2.0 &  --1.4881 & --1.6243 & --1.1549 & AT21, Fig.\  4(g,h)\\
    4.0 &  --0.7443 & --1.6651 & --0.5686 & AT21, Fig.\  4(g,h)\\
   10.0 &  --0.3774 & --1.6105 & --0.4089 & AT21, Fig.\  4(g,h)\\
   20.0 &  --0.8293 & --1.6320 & --0.8539 & AT21, Fig.\  4(g,h)\\
    2.0 &  --0.9788 & --1.6243 & --0.3979 & AT21, Fig.\  4(j,k)\\
    4.0 &  --0.4638 & --1.6021 & --0.3279 & AT21, Fig.\  4(j,k)\\
   10.0 &  --0.2156 & --1.4809 & --0.3279 & AT21, Fig.\  4(j,k)\\
   20.0 &  --0.1963 & --1.5431 & --0.4559 & AT21, Fig.\  4(j,k)\\
    2.0 &  --1.3233 & --1.6243 & --0.9586 & AT21, Fig.\  5(a,b)\\
    4.0 &  --1.0492 & --1.6690 & --0.8239 & AT21, Fig.\  5(a,b)\\
   10.0 &  --0.8239 & --1.8182 & --0.6198 & AT21, Fig.\  5(a,b)\\
   20.0 &  --0.8808 & --1.7804 & --0.6198 & AT21, Fig.\  5(a,b)\\
    2.0 &  --1.5229 & --1.6243 & --1.3979 & AT21, Fig.\  5(d,e)\\
    4.0 &  --1.0231 & --1.6842 & --0.7959 & AT21, Fig.\  5(d,e)\\
   10.0 &  --0.6223 & --1.7626 & --0.5229 & AT21, Fig.\  5(d,e)\\
   20.0 &  --1.0256 & --1.7447 & --0.7696 & AT21, Fig.\  5(d,e)\\
    2.0 &  --1.3979 & --1.6243 & --1.0458 & AT21, Fig.\  5(g,h)\\
    4.0 &  --0.9629 & --1.6690 & --0.6990 & AT21, Fig.\  5(g,h)\\
   10.0 &  --0.5051 & --1.7212 & --0.4202 & AT21, Fig.\  5(g,h)\\
   20.0 &  --0.8751 & --1.6778 & --0.7212 & AT21, Fig.\  5(g,h)\\
    2.0 &  --0.8507 & --1.6133 & --0.3279 & AT21, Fig.\  5(j,k)\\
    4.0 &  --0.6320 & --1.5740 & --0.4437 & AT21, Fig.\  5(j,k)\\
   10.0 &  --0.3010 & --1.5963 & --0.3279 & AT21, Fig.\  5(j,k)\\
   20.0 &  --0.3617 & --1.5624 & --0.3279 & AT21, Fig.\  5(j,k)\\
    2.0 &  --1.3413 & --1.6189 & --0.9586 & AT21, Fig.\  6(a,b)\\
    4.0 &  --1.0872 & --1.6612 & --0.6778 & AT21, Fig.\  6(a,b)\\
   10.0 &  --1.1220 & --1.8094 & --0.6383 & AT21, Fig.\  6(a,b)\\
   20.0 &  --1.2900 & --1.7917 & --0.7212 & AT21, Fig.\  6(a,b)\\
    2.0 &  --1.4881 & --1.6243 & --1.3979 & AT21, Fig.\  6(d,e)\\
    4.0 &  --1.0414 & --1.6612 & --0.6576 & AT21, Fig.\  6(d,e)\\
   10.0 &  --1.0189 & --1.7913 & --0.5528 & AT21, Fig.\  6(d,e)\\
   20.0 &  --1.4337 & --1.7804 & --0.8861 & AT21, Fig.\  6(d,e)\\
    2.0 &  --1.3869 & --1.6133 & --1.0969 & AT21, Fig.\  6(g,h)\\
    4.0 &  --1.0170 & --1.6368 & --0.5850 & AT21, Fig.\  6(g,h)\\
   10.0 &  --0.7867 & --1.7320 & --0.4559 & AT21, Fig.\  6(g,h)\\
   20.0 &  --1.3144 & --1.7192 & --0.8239 & AT21, Fig.\  6(g,h)\\
    2.0 &  --0.7855 & --1.5786 & --0.3565 & AT21, Fig.\  6(j,k)\\
    4.0 &  --0.5898 & --1.4649 & --0.3279 & AT21, Fig.\  6(j,k)\\
   10.0 &  --0.5283 & --1.5506 & --0.3468 & AT21, Fig.\  6(j,k)\\
   20.0 &  --0.7570 & --1.5017 & --0.4685 & AT21, Fig.\  6(j,k)\\
   1.06 &  --0.7373 &    ---   & --0.9788 & Ma21 \\
    1.1 &  --0.7460 &    ---   & --1.0223 & Ma21 \\
    1.4 &  --0.6990 &    ---   & --1.0458 & Ma21 \\
    2.0 &  --0.6394 &    ---   & --0.9393 & Ma21 \\
    3.0 &  --0.5500 &    ---   & --0.8539 & Ma21 \\
    5.0 &  --0.4903 &    ---   & --0.7447 & Ma21 \\
   11.0 &  --0.4044 &    ---   & --1.0000 & Ma21 \\
   21.0 &  --0.5654 &    ---   & --1.0458 & Ma21 \\
   1.04 &  --0.4179 &    ---   & --1.0458 & Sh18, $\lambda_0 = 0.01$~Mm\\
    1.1 &  --0.5287 &    ---   & --1.1871 & Sh18, $\lambda_0 = 0.01$~Mm\\
    1.3 &  --0.7144 &    ---   & --1.3372 & Sh18, $\lambda_0 = 0.01$~Mm\\
    2.0 &  --0.8794 &    ---   & --1.1871 & Sh18, $\lambda_0 = 0.01$~Mm\\
    4.0 &  --0.7773 &    ---   & --1.0458 & Sh18, $\lambda_0 = 0.01$~Mm\\
   11.0 &  --0.6904 &    ---   & --0.9208 & Sh18, $\lambda_0 = 0.01$~Mm\\
   31.0 &  --0.7595 &    ---   & --1.0969 & Sh18, $\lambda_0 = 0.01$~Mm\\
  101.0 &  --0.7212 &    ---   & --1.2596 & Sh18, $\lambda_0 = 0.01$~Mm\\
   1.04 &  --0.3224 &    ---   & --1.0000 & Sh18, $\lambda_0 = 0.1$~Mm\\
    1.1 &  --0.3625 &    ---   & --1.0969 & Sh18, $\lambda_0 = 0.1$~Mm\\
    1.3 &  --0.4535 &    ---   & --1.1612 & Sh18, $\lambda_0 = 0.1$~Mm\\
    2.0 &  --0.4776 &    ---   & --1.0000 & Sh18, $\lambda_0 = 0.1$~Mm\\
    4.0 &  --0.3726 &    ---   & --0.7696 & Sh18, $\lambda_0 = 0.1$~Mm\\
   11.0 &  --0.2503 &    ---   & --0.5376 & Sh18, $\lambda_0 = 0.1$~Mm\\
   31.0 &  --0.2190 &    ---   & --0.7212 & Sh18, $\lambda_0 = 0.1$~Mm\\
  101.0 &  --0.2048 &    ---   & --1.0223 & Sh18, $\lambda_0 = 0.1$~Mm\\
   1.04 &  --0.1574 &    ---   & --0.9586 & Sh18, $\lambda_0 = 1$~Mm\\
    1.1 &  --0.1630 &    ---   & --1.0132 & Sh18, $\lambda_0 = 1$~Mm\\
    1.3 &  --0.1739 &    ---   & --1.0044 & Sh18, $\lambda_0 = 1$~Mm\\
    2.0 &  --0.1637 &    ---   & --0.7696 & Sh18, $\lambda_0 = 1$~Mm\\
    4.0 &  --0.1314 &    ---   & --0.4318 & Sh18, $\lambda_0 = 1$~Mm\\
   11.0 &  --0.0857 &    ---   & --0.2218 & Sh18, $\lambda_0 = 1$~Mm\\
   31.0 &  --0.0857 &    ---   & --0.3979 & Sh18, $\lambda_0 = 1$~Mm\\
  101.0 &  --0.0610 &    ---   & --0.6990 & Sh18, $\lambda_0 = 1$~Mm\\
   1.04 &  --0.0996 &    ---   & --0.8861 & Sh18, $\lambda_0 = 10$~Mm\\
    1.1 &  --0.0996 &    ---   & --0.9208 & Sh18, $\lambda_0 = 10$~Mm\\
    1.3 &  --0.1146 &    ---   & --0.7212 & Sh18, $\lambda_0 = 10$~Mm\\
    2.0 &  --0.1107 &    ---   & --0.4559 & Sh18, $\lambda_0 = 10$~Mm\\
    4.0 &  --0.0904 &    ---   & --0.2076 & Sh18, $\lambda_0 = 10$~Mm\\
   11.0 &  --0.0526 &    ---   &   0.0000 & Sh18, $\lambda_0 = 10$~Mm\\
   31.0 &  --0.0526 &    ---   & --0.1135 & Sh18, $\lambda_0 = 10$~Mm\\
  101.0 &  --0.0701 &    ---   & --0.3468 & Sh18, $\lambda_0 = 10$~Mm\\
   1.05 &  --1.3671 &    ---   & --1.8697 & Sh21 \\
    2.0 &  --1.4881 &    ---   & --1.4815 & Sh21 \\
   2.54 &  --1.3777 &    ---   & --1.3279 & Sh21 \\
    3.0 &  --1.3064 &    ---   & --1.1805 & Sh21 \\
    3.7 &  --1.1928 &    ---   & --1.0000 & Sh21 \\
    5.0 &  --1.0394 &    ---   & --0.8239 & Sh21 \\
   10.0 &  --0.9092 &    ---   & --0.7670 & Sh21 \\
   20.0 &  --0.9126 &    ---   & --0.7959 & Sh21 \\
   30.0 &  --0.9062 &    ---   & --0.9245 & Sh21 \\
   39.5 &  --1.0000 &    ---   & --0.9547 & Sh21 \\
\enddata
\end{deluxetable}

\end{document}